\documentclass[%
 reprint,
 amsmath,amssymb,
 aps,
]{revtex4-1}

\usepackage{aas_macros}
\usepackage{graphicx}
\usepackage{dcolumn,multirow}
\usepackage{bm}
\usepackage{color}
\usepackage{verbatim}
\usepackage{hyperref}
\usepackage[normalem]{ulem}

\begin{document}

\preprint{APS/123-QED}

\title{There is No Missing Satellites Problem}

\author{Stacy Y. Kim$^{1,2}$} \email{kim.4905@osu.edu} 
\author{Annika H. G. Peter$^{1,2,3}$}
\author{Jonathan R. Hargis$^{4}$}
\affiliation{
$^1$Department of Astronomy, The Ohio State University, 140 W. 18th Ave., Columbus, OH 43210, USA
}
\affiliation{
$^2$Center for Cosmology and AstroParticle Physics, The Ohio State University, 191 W. Woodruff Ave., Columbus, OH 43210, USA
}
\affiliation{
$^3$Department of Physics, The Ohio State University, 191 W. Woodruff Ave., Columbus, OH 43210, USA
}
\affiliation{
$^4$Space Telescope Science Institute, 3700 San Martin Drive, Baltimore, MD 21218, USA
}

\date{\today}

\begin{abstract}
A critical challenge to the cold dark matter (CDM) paradigm is that there are fewer satellites observed around the Milky Way than found in simulations of dark matter substructure.  We show that there is a match between the observed satellite counts corrected by the detection efficiency of the Sloan Digital Sky Survey (for luminosities $L \gtrsim$ 340 L$_\odot$) and the number of luminous satellites predicted by CDM, assuming an empirical relation between stellar mass and halo mass.  The ``missing satellites problem'', cast in terms of number counts, is thus solved.  We also show that warm dark matter models with a thermal relic mass smaller than 4 keV are in tension with satellite counts, putting pressure on the sterile neutrino interpretation of recent X-ray observations.  Importantly, the total number of Milky Way satellites depends sensitively on the spatial distribution of satellites, possibly leading to a ``too many satellites" problem.  Measurements of completely dark halos below $10^8$ M$_\odot$, achievable with substructure lensing and stellar stream perturbations, are the next frontier for tests of CDM.


\end{abstract}

\pacs{Valid PACS appear here}
\maketitle



\section{\label{sec:intro}Introduction}

One outstanding problem for the cold dark matter (CDM) paradigm is the missing satellites problem (MSP).  When originally formulated, the MSP highlighted the discrepancy between the number of satellites predicted in CDM simulations, numbering in the 100s, and observed in the Milky Way (MW), numbering $\sim$10 \cite{1993MNRAS.264..201K, 1999ApJ...522...82K, 1999ApJ...524L..19M}.  Since then, increasingly sensitive surveys have pushed the observed satellite count to $\sim$50 (e.g., Ref.~\cite{drlica-wagner2015,Bechtol:2015cbp,Koposov:2015cua}).  Simultaneously, however, improved resolution in numerical simulations has also increased the number of predicted satellites (e.g., \cite{2008MNRAS.391.1685S}).

A crucial step towards resolving the MSP is to correct for those satellites that have not yet been detected.  Only a fraction of the MW's virial volume has been surveyed \cite{2017arXiv170804247N}.  The Sloan Digital Sky Survey (SDSS), by which ultra-faint dwarfs with luminosities as low as 340 L$_\odot$ (Segue I) were discovered, covered only about a third of the sky.  For the faintest dwarfs, SDSS was complete to $\sim$10\% of the MW's virial radius \cite{2008ApJ...686..279K,2009AJ....137..450W}.  The observed count is thus a lower bound on the luminous MW satellite population.  Completeness corrections must be applied to derive the total number of luminous MW satellites.

Fully resolving the MSP requires that the completeness-corrected galaxy count match the predicted \emph{luminous} satellite abundance.  This depends on the physics of an additional key component: baryons.  There is growing evidence that not all dark matter subhalos host an observable galaxy.  Galaxy evolution models \cite{2000ApJ...539..517B} and star-formation histories of ultra-faint dwarfs \cite{2014ApJ...796...91B} indicate that feedback processes and reionization prevent star formation.  In fact, subhalos below $\sim$10$^9$ M$_\odot$ are inefficient in forming a luminous component \cite{2014MNRAS.442.1396W, 2014ApJ...792...99S}.  In CDM, most MW subhalos are dark.

In this work, we compare completeness corrections of the observed MW luminous galaxy population to theoretical predictions for the luminous galaxy population.  We use an analytic approach to highlight specific physics, and provide a roadmap for future MW-based DM constraints.  Our completeness correction is inspired by Refs.~\cite{2008ApJ...688..277T, 2014ApJ...795L..13H, 2016arXiv161207834J, 2017arXiv170804247N}, which used simulations or Bayesian techniques to estimate that the MW hosts hundreds of luminous satellites.  We calculate the total number of luminous galaxies down to 340 L$_\odot$ based on the satellites observed by SDSS.  For comparison, we predict the number of luminous satellites expected in CDM based on empirical scaling relations between halos and galaxies.

Successful dark matter models cannot produce just enough dark matter subhalos to match the corrected galaxy count---they must produce enough \emph{luminous} galaxies.  This places stringent constraints on warm dark matter (WDM) and sterile neutrino models, competitive with Lyman-$\alpha$ forest constraints \cite{Irsic:2017ixq}.

Successful galaxy formation models must produce enough luminous galaxies to match the corrected galaxy count.  This has implications for the mass threshold for the subhalos that host the faintest galaxies, the redshift of reionization, and the tidal stripping of subhalos.


\section{\label{sec:corrections}Completeness Corrections}

The total number of luminous satellites within the MW virial radius ($R_\text{vir}$ = 300 kpc) can be extrapolated from the number of observed satellites by calculating the correction factor $c$ that converts
\begin{equation}
N_\text{tot} = c(\Phi) N_\text{obs}, \label{eqn:c0}
\end{equation}
where $\Phi$ represents the set of parameters the correction depends on.  This includes the survey area, survey sensitivity, and the spatial distribution of satellites.

Recasting in luminosities $L$, and given either a continuous observed luminosity function $dN_\text{obs}/dL$, or set of $N_\text{obs}$ satellites we can express Eq. \ref{eqn:c0} as
\begin{equation}
N_\text{tot} = \int c(L) \frac{dN_\text{obs}}{dL} dL ~ \approx ~ \sum_{i=1}^{N_\text{obs}} c(L_i),
\end{equation}
i.e. we integrate over the luminosity function or sum together the correction for each observed satellite.  The correction is
\begin{equation}
c(L) = \frac{\int_{V_\text{vir}} n(\mathbf{r}) ~ d\mathbf{r}}{\int_{V_\text{obs}(L)} n(\mathbf{r}) ~ d\mathbf{r}} \label{eqn:cM}
\end{equation}
where $n(\mathbf{r})$ is the 3D satellite distribution, $V_\text{vir}$ the MW virial volume, and $V_\text{obs}(L)$ the volume over which a satellite of luminosity $L$ has been surveyed.  Note that the normalization to the spatial distribution cancels, and thus the correction depends only on the shape of the spatial distribution function---not the absolute number of satellites.  Eq. \ref{eqn:cM} naturally accounts for anisotropies in the satellite distribution.

Although there are hints that the luminous satellite distribution is anisotropic \cite{1982Obs...102..202L, 1994ApJ...431L..17M, 2005MNRAS.363..146L, 2008ApJ...688..277T, drlica-wagner2015, bechtol2015, koposov2015, 2017arXiv170804247N}, we assume it is sufficiently isotropic 
to be separable. The correction factor is thus
\begin{equation}
c(L) = c_r(L) ~ c_\Omega(L)
\end{equation}
where $c_r$ and $c_\Omega$ are the radial and angular corrections, respectively, and
\begin{equation}
c_r(L) = \frac{\int_0^{R_\text{vir}} \frac{dN}{dr} ~ dr}{\int_0^{r_c(L)} \frac{dN}{dr} ~ dr}
\hspace{0.5 cm} \text{and} \hspace{0.5 cm}
c_\Omega = \frac{4 \pi}{\int_0^{\Omega_c} d\Omega}
\end{equation}
and $r_c(L)$ is the radius out to which a survey covering an area $\Omega_c$ of the sky is complete for a galaxy with luminosity $L$.  To predict the number of satellites out to a given detection limit for other surveys based on counts from an earlier survey like SDSS, one can replace $R_\text{vir}$ with the radius out to which those surveys are complete.  For SDSS, we adopt the completeness radius derived by Ref. \cite{2009AJ....137..450W}, for which
\begin{equation}
r_c(L) = 15.7 \text{ kpc } \left( \frac{L}{100~\text{L}_\odot} \right)^{0.51}.
\end{equation}
The angular correction dominates for the brightest galaxies, while the radial correction dominates for the faintest galaxies.  The turnover between the two occurs at roughly L = 500 - 2000 L$_\odot$ (depending on the satellite distribution), and rapidly becomes large at lower luminosities.


\begin{figure}
\includegraphics[width=0.5\textwidth]{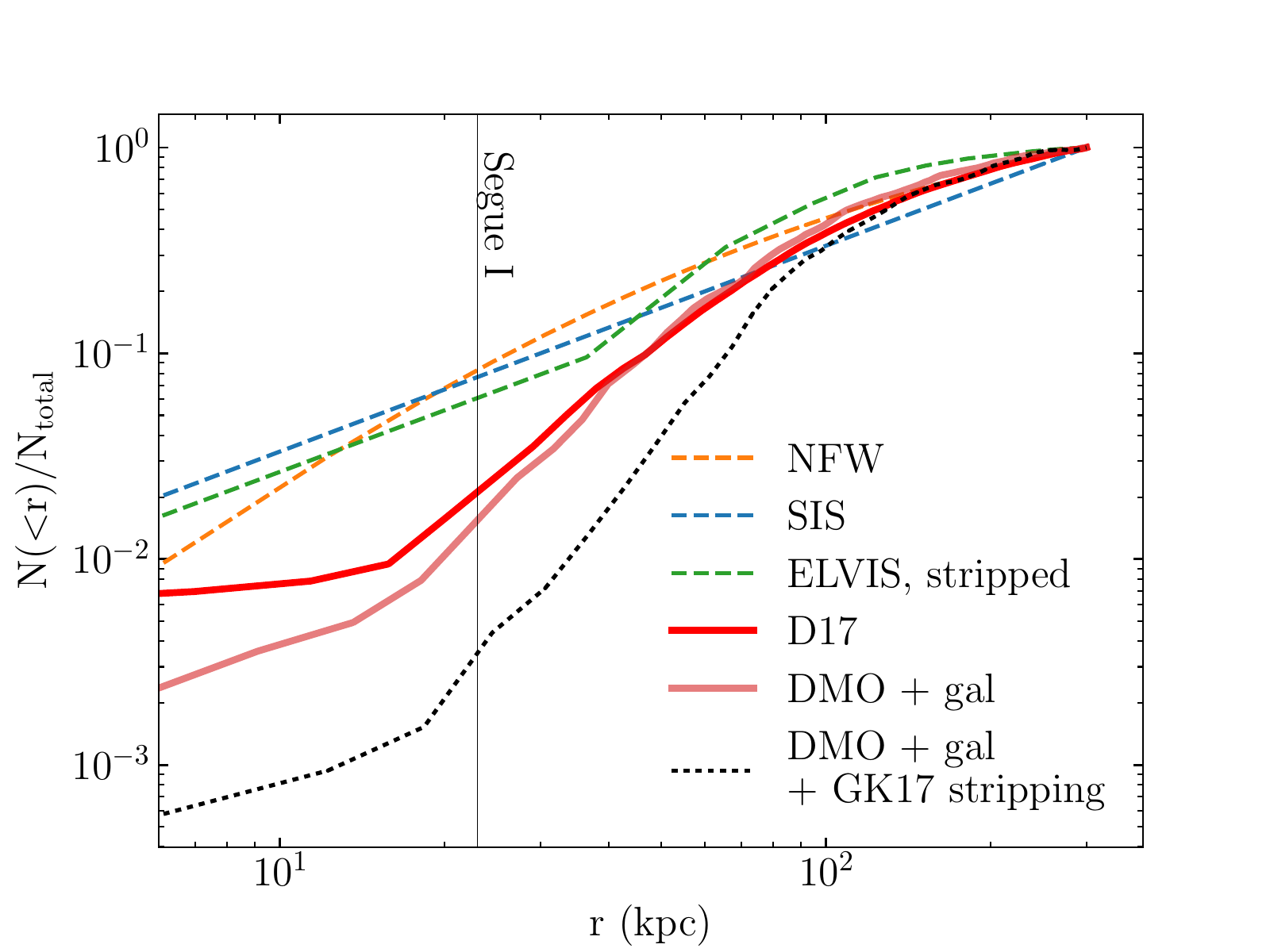}
\caption{Normalized radial distributions.  The cumulative number of satellites within radius $r$, normalized to the total number of satellites at $R_\text{vir}$ = 300 kpc, is shown.  Distributions marked by dashed lines are expected when satellites survive extreme tidal stripping. The solid red lines are our fiducial radial distributions, matching the MW classical satellites.  The dotted black line depicts the latter distribution depleted by Ref.~\cite{2017MNRAS.471.1709G}'s tidal stripping model.  See text for details.}
\label{fig:profiles}
\end{figure}

The radial distribution of luminous satellites, which is highly uncertain, has a significant impact on the corrected galaxy count.  Well-motivated radial profiles from the literature, spanning the range of uncertainty on tidal stripping and subhalo-galaxy identification, are shown in Fig.~\ref{fig:profiles}. The centrally concentrated NFW (with concentration $c_\text{-2}$ = 9) \cite{1996ApJ...462..563N} and SIS (singular isothermal sphere) \cite{2011ApJ...731...44N} models correspond to the smooth dark matter distribution of the host.  These profiles include satellites that are severely tidally stripped \cite{2016MNRAS.457.1208H, 2017arXiv170804247N}, which may be considered destroyed in other contexts (or unresolved in numerical simulations). The light red line is representative of distributions generated from dark matter only (DMO) simulations, but with assumptions on which subhalos host galaxies \cite{2014ApJ...795L..13H, 2017MNRAS.471.4894D}.  The black dotted line shows how tidal stripping by a baryonic disk reduces the number of satellites close to the center of the MW as in Ref.~\cite{2017MNRAS.471.1709G} (hereafter GK17; see also \cite{2010ApJ...709.1138D}).  In contrast, the most severely stripped halos in Ref.~\cite{2014MNRAS.438.2578G} are shown in green.  This corresponds to the hypothesis that the SDSS satellites are highly stripped remnants of larger galaxies.  For our fiducial distribution, we adopt that derived in Ref.~\cite{2017MNRAS.471.4894D} (hereafter D17; shown by the dark red line), which matches the distribution of observed classical MW satellites.

\begin{figure*}
\centerline{
\includegraphics[scale=0.6]{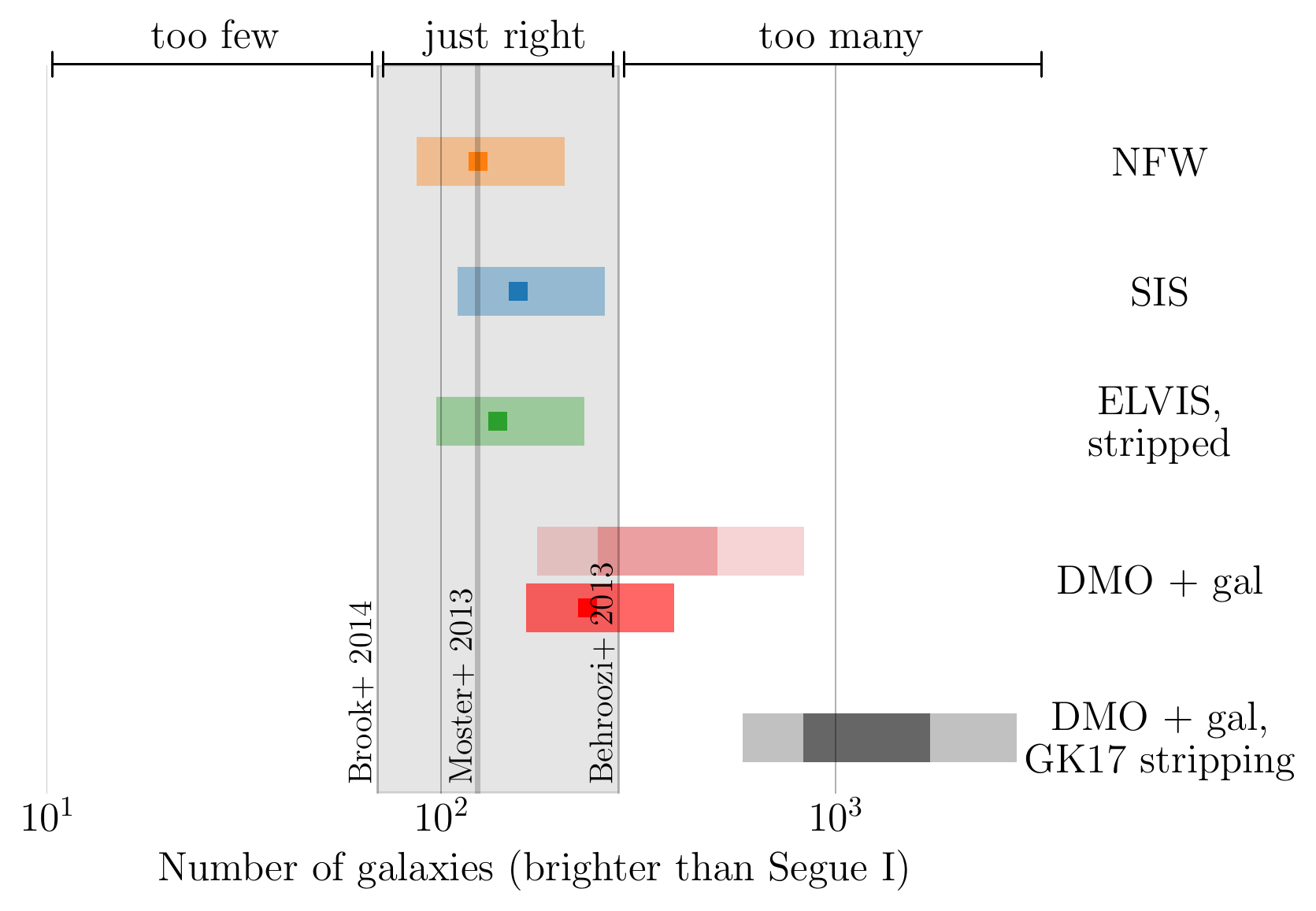}
\includegraphics[scale=0.6]{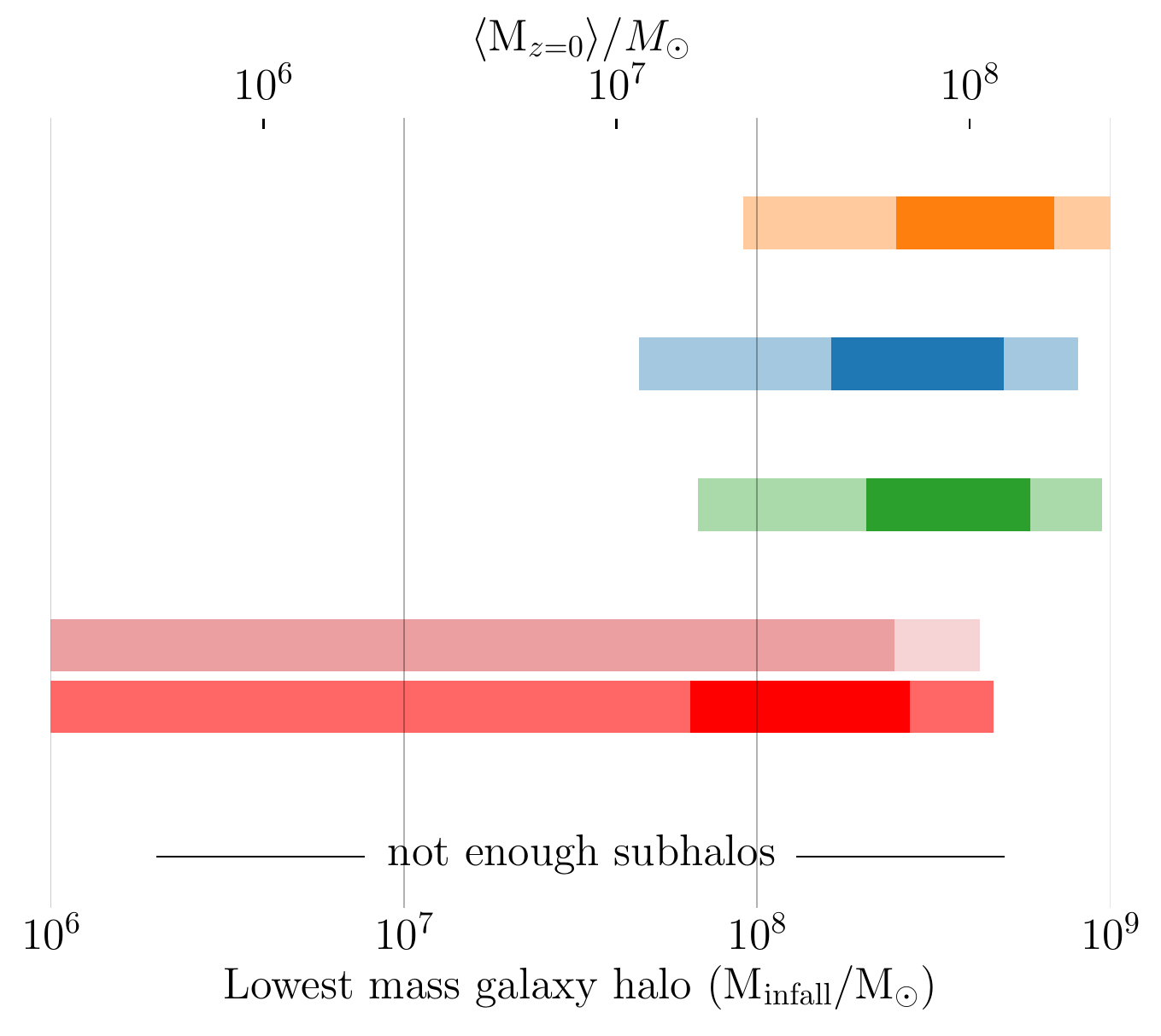}}
\caption{The number of completeness corrected luminous satellites (left) and the infall mass of the lowest mass subhalo hosting a $L > 340 L_\odot$ galaxy (right). Color match those in Fig.~\ref{fig:profiles}; the dark red denotes results based on D17.  The light colored bands denote the uncertainty due to anisotropy (based on \cite{2008ApJ...688..277T}).  Left: The gray-shaded region shows the predicted number of luminous satellites expected for the MW based on the calculation described in Sec. \ref{sec:constraints}.  If the completeness-corrected count falls within these bounds, there is no MSP. Right: The width of the dark bands is set by the uncertainty on the MW mass, (1-2) $\times 10^{12}$ M$_\odot$.  The light bands denote uncertainties due to anisotropy.  The bottom axis shows the subhalo mass at infall, and the top axis shows the average corresponding subhalo mass today \cite{2005MNRAS.356.1233V}.}
\label{fig:ntot}
\end{figure*}

We correct the number of galaxies observed through SDSS Data Release 8 (DR8) with these radial profiles.  Our results are shown in the left panel of Fig.~\ref{fig:ntot} and are listed in Tab.~\ref{tab:counts}.  The width of the bars denote the uncertainty due to anisotropy, as measured in \cite{2008ApJ...688..277T} (see supplementary material for a detailed discussion of anisotropies).  In agreement with Refs. \cite{2014ApJ...795L..13H, 2017arXiv170804247N}, we find that the correction is insensitive to the MW halo mass.

Radial distributions corresponding to the hypothesis that satellites survive extreme tidal stripping (NFW and SIS) are more centrally concentrated, resulting in smaller corrected counts.  Accounting for the effects of tidal stripping due to the presence of a baryonic disk as predicted by \cite{2017MNRAS.471.1709G} produce the largest corrections.

These results are a lower limit to the number of luminous satellites of the MW.  We have not included the satellites of the Large Magellanic Cloud, dwarfs with surface brightnesses $\mu \le$ 30 mag arcsec$^{-2}$ (``stealth galaxies", e.g. \cite{2010ApJ...717.1043B, 2016MNRAS.459.2370T}), which are below the detection limit of SDSS DR8, although they have been found by new surveys \cite{drlica-wagner2015,2016MNRAS.459.2370T}, and dwarfs with luminosities below Segue I's.  Segue I itself accounts for $\sim$40\% of the correction; accounting for even fainter galaxies will increase the total number significantly.  The inferred luminosity function of satellites is shown in the supplementary material.

\begin{table}
\caption{Completeness corrected satellite counts}
\label{tab:counts}
\begin{tabular}[b]{l c c c} \hline
&\multicolumn{3}{c}{Predictions} \\
distribution & all sky & DES & LSST Year 1 \\ \hline
NFW & 124 & 11 & 56 \\
SIS & 157 & 13 & 69 \\
ELVIS, stripped & 139 & 13 & 65 \\
D17 & 235 & 18 & 102 \\
DMO + gal & 250-503 & 20-28 & 109-198 \\
DMO + gal + GK17 & 830-1740 & 49-69 & 335-614 \\ \hline
\multicolumn{4}{p{\linewidth}}{\footnotesize Predictions for DES, when complete after year 5, and sensitive down to apparent magnitudes $V$ = 24.7; and for LSST after year 1, down to $V$ = 26.} \\
\end{tabular}
\end{table}


\section{\label{sec:constraints}Constraints on galaxy evolution and dark matter models}

The calculations above set only the total number of \emph{luminous} MW satellites that we can infer exists based on the observed dwarfs.  Do the corrected counts imply that the MSP is solved?  We present our fiducial calculation here, and provide details on choices and variants in the supplementary material.

The number of dark matter subhalos hosted by the MW is derived by integrating the CDM mass function, which follows the form
\begin{equation}
\frac{dN}{dM} = K_0 \left( \frac{M}{M_\odot} \right)^{-\alpha} \frac{M_\text{host}}{M_\odot}.
\end{equation}
where $M$ denotes the mass of a subhalo at infall.  The mass function based on present day (e.g. $z$ = 0) subhalo masses is lower due to tidal stripping ($M(z=0) < M$).  We adopt $K_0 = 1.88 \times 10^{-3}$ M$_\odot^{-1}$ and $\alpha = 1.87$ as in D17.  The total number of subhalos above a threshold $M_\text{min}$ is thus
\begin{equation}
N_\text{sub} = \int^{M_\text{host}}_{M_\text{min}} \frac{dN}{dM} ~ dM.
\end{equation}
Not all subhalos are believed to host galaxies \cite{2016MNRAS.456...85S, 2017arXiv170503018S, 2002MNRAS.333..177B, 2002ApJ...572L..23S}.  Given the fraction of subhalos of a given mass that host a luminous galaxy, $f_\text{lum}(M)$, we can derive the total number of luminous galaxies
\begin{equation} \label{eq:reionization}
N_\text{gal} = \int^{M_\text{host}}_{M_\text{min}} \frac{dN}{dM} f_\text{lum}(M) dM.
\end{equation}
The luminous fraction is a strong function of reionization redshift. $z_\text{re}$, and the survival criteria \cite{2000ApJ...542..535G}. We adopt the relation by D17 (see their Fig. 3), which assumes $z_\text{re}$ = 9.3 and a generous baryon survival criterion, which requires $v_\text{max}$ = 9.5 km/s (the peak of the circular velocity curve) at $z < z_\text{re}$ and $v_\text{peak} = 23.5$ km/s at $z > z_\text{re}$.  Adopting a less generous criterion to match other work in the literature \cite{2013MNRAS.432.1989S, 2014ApJ...786...87B, 2015ApJ...807L..12O} drops the predicted number of luminous satellites by a factor 2---our results thus represent an upper bound.

For comparison with our completeness correction, which only includes galaxies brighter than Segue I, we adopt $M_\text{min}$ = $M_\text{SegI}$.  We derive its total stellar mass by assuming a stellar mass-to-light ratio of 2 (i.e. $M_*^\text{SegI} = 680$ M$_\odot$, expected for an ancient metal-poor stellar population with a Kroupa initial mass function \cite{kroupa2001, martin2008}.  To derive Segue I's halo mass, we make use of the fact that a galaxy's stellar mass is empirically tightly correlated with halo mass \citep{2004MNRAS.353..189V, 2004ApJ...609...35K}, a relation known as the stellar-mass--halo-mass (SMHM) relation.  SMHM relations have only been calibrated for stellar masses greater than $M_*\sim 10^8M_\odot$ \cite{2013ApJ...770...57B, 2013MNRAS.428.3121M}, but hydrodynamic simulations indicate that extrapolations to low masses are reasonable \cite{2017arXiv170506286M}.  We adopt three SMHM relations that capture the diversity of SMHM relations and their scatter \citep{2014ApJ...784L..14B, 2013MNRAS.428.3121M, 2013ApJ...770...57B}, which gives a large range for $M_\text{SegI}$ = 8$\times$10$^6$ to 7$\times$10$^8$ M$_\odot$.  The SMHM of Moster et al. \cite{2013MNRAS.428.3121M} best matches hydrodynamic simulations of isolated galaxies \cite{2014MNRAS.445..581H,2017arXiv170506286M}.

\begin{figure}
\includegraphics[width=0.5\textwidth]{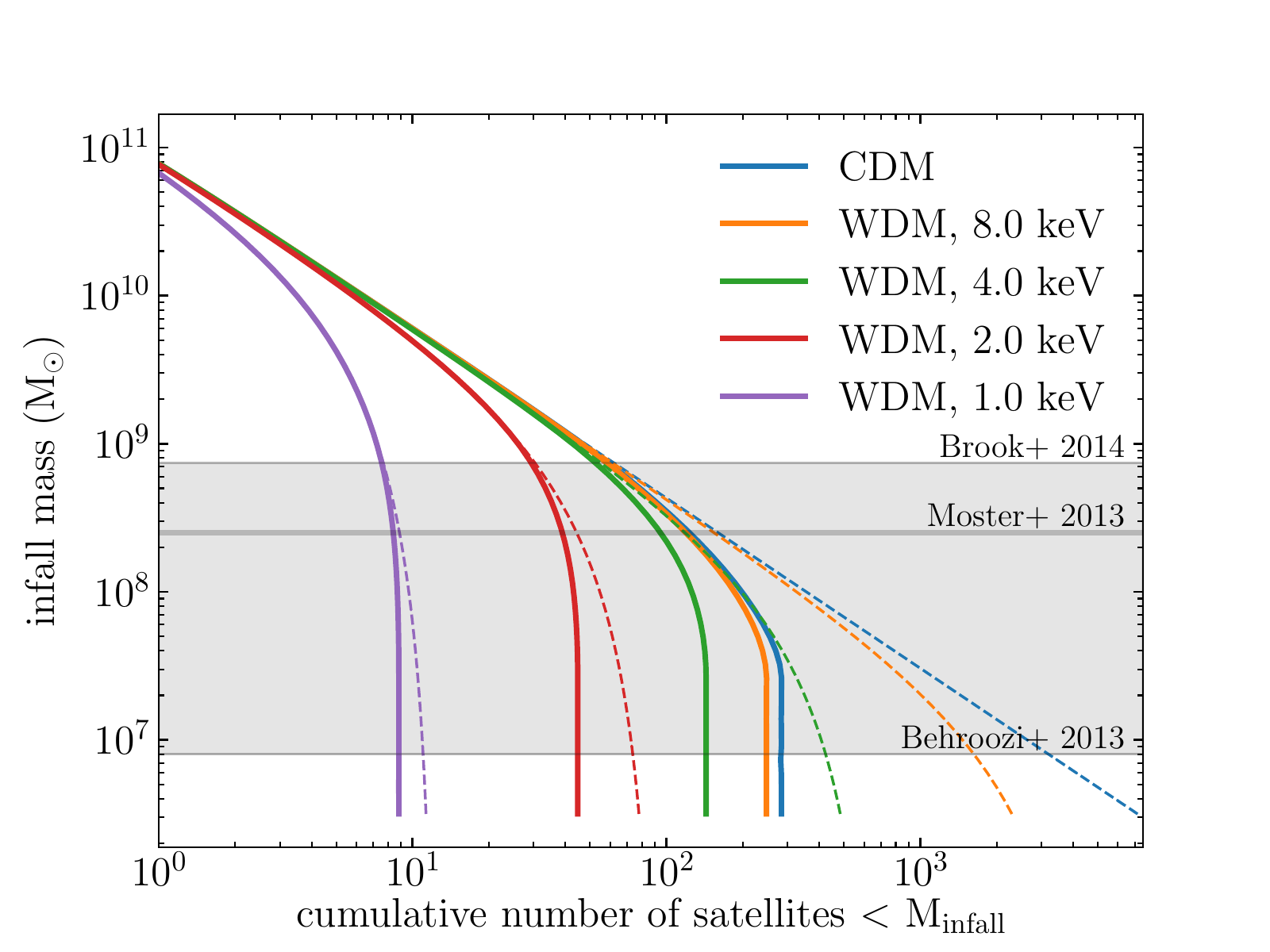}
\caption{The number of luminous (solid) and total (dashed) subhalos allowed by CDM and WDM mass functions for a given lower bound on the lowest mass subhalo, assuming a MW mass of 1.5 $\times$ 10$^{12}$ M$_\odot$.  The number of luminous subhalos is modeled as in D17, assuming a reionization redshift $z_\text{re}$ = 9.3 \cite{2016A&A...596A.108P}.  The grey band shows SHMH predictions for the infall mass of Segue I.}
\label{fig:mf_nsubs}
\end{figure}

The resultant number of subhalos and galaxies assuming a MW mass of 1.5 $\times$ 10$^{12}$ M$_\odot$ is shown in Fig.~\ref{fig:mf_nsubs}.  The solid line denote the number of luminous satellites predicted by CDM; the dashed line shows the total number of subhalos.  Down to $M_\text{min}$ = 10$^7$ M$_\odot$, there exists $\sim$2600 total subhalos and 280 galaxies, implying that only $\sim$10\% of such subhalos are luminous.  The number of luminous satellites down to $M_\text{SegI}$ for our range of SMHM relations is shown by the gray-shaded regions in Fig.~\ref{fig:mf_nsubs} and the left-hand panel of Fig.~\ref{fig:ntot}.  The SMHM relation of Moster et al. \cite{2013MNRAS.428.3121M} predicts $\sim$120 galaxies.

We now have all the tools required to match the theoretical and completeness-corrected galaxy counts.  Our key result is shown in the left-hand plot of Fig. \ref{fig:ntot}.  The number of galaxies more luminous than Segue I predicted in CDM matches the completeness-corrected observations for even the most conservative radial profile models (i.e. lies inside the gray-shaded region).  Moreover, some radial profiles lead to corrected counts that exceed the predicted range.  This is exacerbated if the reionization of the Local Group occurs earlier (see supplementary material).  We call this the ``too many satellites" problem.

These results have implications for galaxy formation theory and dark matter physics.

\emph{Subhalo minimum mass}.  Matching the corrected counts to CDM predictions gives a minimum subhalo mass for galaxies as faint as Segue I, as suggested by Ref.~\cite{2016arXiv161207834J}.  This is shown on the right panel of Fig.~\ref{fig:ntot}.  The bars denote the uncertainty on the lowest mass galaxy halo due to uncertainties on the MW mass, which we allowed to range from  $(1-2)\times$ 10$^{12}$ M$_\odot$.  As the transition from mostly bright to mostly dark subhalos (e.g. $f_\text{lum}$ = 0.5) occurs at $\sim$10$^8$ M$_\odot$ in our reionization model, the lowest mass galaxy is near that mass threshold.  Counts accounting for tidal stripping as in GK17 predict even smaller masses.  The tidal-stripping-induced uncertainty on the completeness correction is the single-biggest driver of uncertainty in the subhalo minimum mass.

\emph{Dark matter model}.  Dark matter models with suppressed matter power spectra \cite{2016PhRvD..93l3527C, 2014MNRAS.442.2487K, 2017PhRvD..95h3015C, 2016MNRAS.461...60L, 2017MNRAS.468.4285L, 2016MNRAS.463.3848B, 2017MNRAS.464.4520B, 2011PhRvD..83d3506P,2017PhRvD..95d3541H, 2014NatPh..10..496S, 2016PhRvD..94d3513S, 2017MNRAS.471.4559M, 2010IJMPD..19.1385K}, such as WDM, must reproduce \emph{at a minimum} the completeness-corrected counts.  We briefly sketch the constraints we can place on WDM models with our corrected counts.  This calculation is not intended to be a rigorous derivation of WDM constraints, but an illustration of possible limits when corrected counts are taken into account.

The radial distribution of WDM satellites closely follows CDM \cite{2014MNRAS.439..300L}, and thus the corrected counts derived above applies.  To obtain the number of luminous satellites predicted by WDM, an identical analysis as in the previous section can be performed with the WDM mass function, which can be obtained from the CDM mass function by multiplying the factor
\begin{equation}
	\frac{dn_\text{WDM}}{dn_\text{CDM}} = \left( 1 + \frac{M_\text{hm}}{M} \right)^{-\beta},
\end{equation}
where $\beta = 1.16$, $M$ is the infall mass, and $M_\textrm{hm}$ is the half-mode mass quantifying the suppression scale of the matter power spectrum \cite{2012MNRAS.424..684S}. We again use D17's reionization cutoff to estimate the number of luminous galaxies.  This is a conservative overestimate, as WDM halos tend to form, and form stars, later than in CDM \cite{2018arXiv180305424B}, and because some of the satellites will be fainter than Segue I.  In Fig.~\ref{fig:mf_nsubs}, we show the number of satellites predicted by WDM mass functions for thermal relic particle masses ranging from 1-8 keV.  If the MW has 120-150 galaxies brighter than Segue I, thermal relics below $\sim$4 keV are ruled out, although this depends on the MW halo mass (\cite{2014MNRAS.442.2487K}, and see supplementary material).  This implies that the 7 keV sterile neutrino is in tension with satellite counts.  More robust limits require the machinery of Ref. \cite{2016arXiv161207834J}, who find a 95\% lower limit of 2.9 keV.


\section{Conclusions}

Since the MSP was first identified, several advances in our understanding of dwarf galaxy evolution have reduced the severity of the missing satellites problem.  Star formation in low-mass halos has been demonstrated to be suppressed by reionization and feedback.  The discovery of many new dwarfs below the luminosity limit of the classical dwarfs have also closed the gap, as has the understanding that completeness corrections for the new dwarfs are large.  In this Letter, we show that such corrections imply that the number of satellite galaxies that inhabit the Milky Way is consistent with the number of luminous satellites predicted by CDM down to halo masses of $\sim$10$^8$ M$_\odot$.  There is thus no missing satellites problem.  If anything, there may be a ``too many satellites" problem.  The major remaining uncertainty is the radial distribution of satellites, stemming from the uncertainty in tidal stripping.  Our result pushes the scale for tests of CDM below $10^8 M_\odot$ in infall mass, or $\sim$10$^7$ M$_\odot$ in present day subhalo mass.  Methods that do not rely on baryonic tracers, like substructure lensing \cite{2014MNRAS.442.2434N, 2016JCAP...11..048H, 2018arXiv180101505V} or stellar stream gaps \cite{2017MNRAS.466..628B}, are required to test the predictions of CDM below this scale.  The implications for dark matter models are significant.  WDM theories equivalent to having thermal relic particle masses below 4 keV are in tension with MW satellite counts.

\section*{Acknowledgments}

We thank John Beacom, Andrew Benson, Matthew Buckley, Francis-Yan Cyr-Racine, Greg Dooley, Anna Nierenberg, Todd Thompson, and David Weinberg for helpful comments.  AHGP acknowledges support from National Science Foundation (NSF) Grant No. AST-1615838. 

\bibliography{no_msp}

\clearpage
\appendix

\section{Model choices}
Our goal with this work is to show that with well-motivated theoretical choices, there is no missing satellites problem for the MW and to give (in particular) particle theorists a roadmap to test their preferred dark matter models with MW satellite counts.  In this section, we describe the choices we made for models in the main text, and discuss how our key results depend on these choices.  If, due to uncertainties, a range of choices are permissible, we generally chose the most conservative option.  We show that despite this approach, we are led to the conclusion that there are no missing satellites.

\subsection{Spatial distribution of satellites and the completeness correction}
As we showed in Sec. II, the completeness correction depends sensitively on assumptions of how luminous satellites are distributed in the halos of their hosts.  We consider the radial and angular distribution of satellites separately.

\vskip 0.1cm
\noindent
\emph{Radial distribution}.
There is considerable uncertainty in the radial distribution of surviving (i.e., not tidally disrupted) subhalos and luminous satellites.  Our choice to use a range of analytic radial distributions for our completeness corrections spans this uncertainty, which stems primarily from whether the innermost subhalo population is faithfully represented in simulations, and whether the distributions of luminous and dark satellites differ.  We discuss these uncertainties below.

There are hints that even high-resolution simulations today systematically underpredict the abundance of the innermost satellites. This is concerning as the completeness correction depends sensitively on them.  If subhalos can survive being stripped of the vast majority of their mass ($\gg 99$\%), their distribution should follow that of the host's smooth halo component \cite{2016MNRAS.457.1208H}.  However, in simulations, subhalos are biased away from the halo centers at fixed subhalo mass \cite{2010AdAst2010E...8K, 2016MNRAS.457.1208H}.  This may be caused by numerical overmerging \cite{vandenBosch:2016hjf, vandenBosch:2017ynq, vandenBosch:2018tyt}, which can underestimate the amount of substructure in halos and make the subhalo population appear less concentrated than it really is.  The major consequence is producing completeness corrections in the direction of a ``too many satellites'' problem.

Semi-analytic modelers are beginning to use an ``orphan galaxy'' (subhalos added to represent unresolved, artificially destroyed subhalos in the simulation) correction to radial distributions \cite{Simha:2016poj, 2017arXiv170804247N}, but the correction needs better calibration on scales relevant to the Milky Way satellite population.  In particular, it is not clear what level of stripping is allowed for us to perceive a satellite as being intact instead of appearing severely tidally disturbed.  In the main text, we thus considered a distribution that matches the cuspy NFW DM distribution found in simulations.  This choice gives the most conservative completeness correction (a la Ref. \cite{2017arXiv170804247N}).  However, it does appear somewhat more concentrated than observed massive satellite populations at moderate redshift \cite{2011ApJ...731...44N}. 

In addition to numerical artifacts, there is a concern that the internal density profile of satellites assumed---which is itself uncertain---affects the satellite abundance and distribution.  Although dark matter-only CDM simulations predict satellites with steeply rising central density cusps, the addition of baryonic (or new dark matter) physics can produce shallower cores.  Several authors find that cores substantially reduce satellite survival, especially toward host centers \cite{Brooks:2012ah, 2017MNRAS.465L..59E}, while others do not \cite{2017MNRAS.471.1709G}. Similar to numerical overmerging, cored models would increase the completeness correction and decrease the predicted CDM satellite abundance, exacerbating a ``too many satellites'' problem.  For our analysis, we thus conservatively assume radial distributions appropriate for cusped satellites.  Furthermore, we note that if cores are generated by baryons, then only the classical satellites are likely to have them; the ultrafaint galaxies do not form enough stars to destroy their cusps \cite{Penarrubia:2012bb, 2017MNRAS.471.3547F}.

Finally, we consider the effect of the baryonic disk on the satellites.  Several studies show that the presence of the Milky Way's disk drastically increases tidal stripping of satellites within a few tens of kpc from the center of the host \cite{2010ApJ...709.1138D, 2017MNRAS.471.1709G, 2017MNRAS.465L..59E}.  The least centrally concentrated distribution we considered combines the D17 selection of luminous subhalos with a GK17-based disk stripping cut.  This may be unrealistically centrally depleted, given that it results in a severe too many satellites problem.  It is possible that numerical overmerging is significant. Nevertheless, it is well-motivated by simulations and represents the model with the largest completeness correction.

Aside from concerns regarding the abundance and distribution of subhalos, there is a question of which of the subhalos are luminous.  There are on-going discussions as to how the subhalo distribution is modified when considering only \emph{luminous} satellites (e.g. \cite{2014ApJ...795L..13H,2017MNRAS.471.4894D}).  For a given infall halo mass, luminous subhalos near the transition from $f_\text{lum} \approx 1$ to $f_\text{lum} \approx 0$ were preferentially formed and accreted onto the host early.  D17 find that luminous satellites are hence more centrally concentrated than the subhalo population as a whole, given a subhalo mass threshold, and that their fiducial satellite radial distribution is a good match to the classical Milky Way satellites.  We thus adopt the radial distributions of Ref. \cite{2014ApJ...795L..13H} and D17 as our fiducial radial distributions in this work.

The distributions that we considered thus represent a realistic range, spanning the range of theoretical uncertainty in satellite survival.  

\vskip 0.1cm
\noindent
\emph{Angular distribution and anisotropy}.
There are four potential causes for satellite anisotropy which we consider as we calculate our completeness correction.

First, the Sun is offset from the center of the MW.  However, Ref. \cite{2014ApJ...795L..13H} found that accounting for the fact that we observe the satellite distribution at 8 kpc from Galactic Center does not affect completeness corrections, and presumably impose no significant anisotropies.

Second, there is intrinsic anisotropy in simulations of Milky Way-mass systems, originating from the anisotropic assembly history of halos \cite{2008ApJ...688..277T,2014ApJ...795L..13H}.  We account for this uncertainty in the main text.  In detail, \cite{2008ApJ...688..277T} found that for SDSS DR5, which covered $\sim$1/5 of the sky, the area (or containment) correction ranged from $c_\Omega$ = 3.5 to 8.3, 60-170\% of the nominal survey area coverage.  We conservatively assume the same level of variation due to anisotropy for DR8, for which $c_\Omega \sim$ 1/3.

Third, there are claims of a satellite plane around the Milky Way (e.g. \cite{1976RGOB..182..241K, 2007MNRAS.374.1125M,2018MPLA...3330004P}).  Some simulations show that the distribution of subhalos are not spherically symmetric \citep{2005ApJ...629..219Z, 2005A&A...437..383K}.  If the SDSS dwarfs were part of a satellite plane, the completeness correction would be smaller than we indicated in the main text. However, there is recent evidence that ultrafaint dwarfs do not live in planes. The 3D velocities of the SDSS dwarfs obtained by Gaia indicate that the orbits are substantially different from those of the classical satellites, and that the angular momentum vectors of their orbits are not well aligned with that of the plane of satellites \cite{2018arXiv180410230S, 2018arXiv180507350F, 2018arXiv180500908F, 2018arXiv180501839M}.  Moreover, the newly discovered Milky Way stellar streams in the DES footprint, which may be the remnants of ultrafaint galaxies (the best validated stream-galaxy connection is Tucana III's) have a wide distribution of angular momentum vector orientations \cite{2018arXiv180103097S, 2018arXiv180407761L}.  Thus, we conclude that the orbits of the known ultrafaint galaxies (including those that are disrupted into streams) are inconsistent with a plane-of-satellites origin.

Finally, we consider the possibility that the LMC satellites can add an anisotropic contribution to the satellite distribution.  As stated in the text, however, we are only concerned with galaxies that are satellites of the Milky Way, and not satellites of satellites (or in simulations, sub-subhalos).  D17 showed that including LMC satellites causes only a 10\% change in the Milky Way's satellite count.  However, there are claims that the LMC satellites can account for upwards of 30\% of the MW satellites (e.g. \cite{2016MNRAS.461.2212J}).  This is likely too high an estimate.  Eqn. 7 in the main text shows that the number of subhalos scales with the mass of the host; given that the LMC's dark matter halos is about 1/10 the mass of the MW's, it likely contributes at most 10\% of the MW's satellites (and likely less, given that a larger fraction of the LMC satellites should be dark relative to the MW).  It is possible that the radial distribution adopted or the LMC satellites in Ref. \cite{2016MNRAS.461.2212J}, which is not centrally concentrated, results in an overly large estimate of the satellite population.

In conclusion, there is substantial evidence that the distribution of ultrafaint galaxies is largely isotropic.  Any anisotropies are a product of the intrinsic anisotropy in the orientation of accreted satellites, an observation borne out in high-resolution simulations of Milky Way-mass halos.  We include this source of anisotropy in our completeness correction uncertainty budget.

\subsection{CDM halo mass function and connection to Milky Way mass}

For the CDM subhalo mass function upon which our luminous satellite prediction is built, we take the average mass function from the \emph{Caterpillar} CDM-only simulation of MW analogs \cite{Griffen:2016ayh,2017MNRAS.471.4894D}.  We checked that this mass function agrees within 10\% of that derived for the ELVIS simulations of the Local Volume \cite{2014MNRAS.438.2578G}.

There are two minor uncertainties in the subhalo mass function.  At fixed subhalo mass, the halo-to-halo scatter in the subhalo mass function is well-described by a negative binomial distribution with a width of about 15\%, which is small \cite{BoylanKolchin:2009an, 2014MNRAS.438.2578G, 2017MNRAS.472.1060D}. Another small uncertainty on the subhalo mass function relates to the presence of baryons.  Simulations show that baryonic physics suppresses the low-mass subhalo mass function by $\mathcal{O}(10\%)$, from a combination of suppressed mass accretion and enhanced stripping by the MW disk \cite{2016MNRAS.456...85S, 2017MNRAS.471.1709G, 2017MNRAS.465L..59E}.  We consider these to be small effects for CDM satellite predictions.

The largest uncertainty in the subhalo mass function comes from the uncertainty in the mass of the Milky Way halo.  The subhalo abundance is directly proportional to the mass of the MW (e.g. Eq. 7).  While we have adopted M$_\text{MW} = 1.5 \times 10^{12}$ M$_\odot$, current estimates of the Milky Way mass range from $0.5-2 \times 10^{12}$ M$_\odot$.  However, masses on the lower end of this range cause the existence of the Milky Way's large dwarf companions, the Large and Small Magellanic Clouds, to become increasingly problematic \cite{2011ApJ...738..102T, 2011ApJ...743..117B}.  Such a low mass is also inconsistent with expectations from SMHM relations \cite{2013MNRAS.428.3121M, 2013ApJ...770...57B}. We conservatively chose a mass toward the upper end of the likely range---resulting in a larger number of satellites predicted from simulations. The satellite abundance drops from 124 (160) luminous satellites (all subhalos) for a MW mass of $1.5 \times 10^{12}$ M$_\odot$ to 83 (108) for a MW mass of $10^{12}$ M$_\odot$, and 41 (53) for a MW mass of $0.5 \times 10^{12}$ M$_\odot$, assuming an infall subhalo mass $M_\text{Seg I}$ given by the SMHM relation of Ref. \cite{2013MNRAS.428.3121M}.  Adopting a smaller MW mass than we assumed goes even further in the direction of a no missing satellites problem and pushes us well into the regime of a ``too many satellites'' problem.

\subsection{Gas cooling, reionization, and the luminous fraction of subhalos}

Given a subhalo mass function, which subhalos host observable galaxies?  We here discuss the relevant pieces of galaxy formation physics that determine which subhalos are luminous.

What is the smallest halo that might host a galaxy?  Stars form from cold, dense gas, and thus the benchmark lowest-mass galaxy halos are assumed to have $T_\mathrm{vir} = 10^4$ K, the temperature above which atomic hydrogen cooling becomes efficient, and at which halos can retain gas after supernova feedback.  This corresponds to a halo mass of around $10^8 M_\odot$ before reionization \cite{1996ApJ...465..608T, 2017ApJ...848...85J, 2016ApJ...831..204R}.  Recently, simulations have shown that metal-line cooling can be effective in some halos down to a mass of $M_\mathrm{vir} \sim 10^7 M_\odot$ unless the local radiation field is strong \cite{Simpson:2012ax, wise2014, 2015ApJ...807L..12O, 2015ApJ...807..154B, 2016ApJ...833...84X}.  However, although some galaxies can form in such small halos, simulations show that most $10^7 M_\odot$ halos at $z\sim 10$ do not contain stars because of feedback from other nearby ionizing sources.  More typically, Refs. \cite{2015ApJ...807L..12O, 2016ApJ...833...84X} find that about half of $M_\mathrm{vir} = 10^8 M_\odot$ (atomic cooling) halos contain stars. 

Star formation is universally suppressed in small halos by the end of the reionization epoch.  Ionizing photons from small galaxies and Pop III stars photoevaporate gas out of existing small halos \cite{1999ApJ...523...54B}, preventing small halos to accrete gas, much less cool it.  The scale below which halos possess 50\% or less of the universal baryon fraction of gas is called the ``filtering scale'' \cite{2000ApJ...542..535G}.  The filtering scale is typically quantified in terms of $v_\mathrm{vir}$ or a related characteristic velocity.  At reionization ($z\sim 10$), the filtering velocity scale is 30-50 km/s \cite{1996ApJ...465..608T, 2000ApJ...542..535G, 2008MNRAS.390..920O, Dawoodbhoy:2018rzc}, corresponding to virial masses $M_\mathrm{vir}(z) > 10^9 M_\odot$.  It was recognized early that this threshold was sufficiently low enough that it could dramatically affect the observable MW satellite population \cite{2000ApJ...542..535G, 2002MNRAS.333..156B, 2002MNRAS.333..177B, 2002ApJ...572L..23S}.  Only a small fraction of halos below this threshold form stars. After reionization, only significantly more massive halos admit star formation.  

In semi-analytic models like ours, it is common to translate these findings into a parameterized two-step model to determine the luminous fraction.  We use a threshold parameterized in terms of $v_\mathrm{max}$, the peak of the $\sqrt{GM(r)/r}$ curve which is typically $\sim 10\%$ higher than $v_\mathrm{vir}$, as the condition for the two-step prescription we adopted from Ref. \cite{2017MNRAS.471.4894D}.  We choose this parameterization as $v_\text{max}$ tends only rises modestly after the initial halo formation epoch.  It also tracks well with any temperature-related conditions.  We require halos to achieve a minimum $v_\mathrm{max}$ = 9.5 km/s (corresponding to a halo mass of $10^7M_\odot$ at $z=10$) before reionization to have a luminous component, or $v_\mathrm{max}$ = 23.5 km/s after reionization.  Our pre-reionization threshold is lower than the atomic hydrogen cooling halo limit, in accordance with Refs. \cite{2015ApJ...807L..12O, 2016ApJ...833...84X}.  It is also a good match to Ref. \cite{2014MNRAS.437..959B}.  We set the post-reionization threshold to be below the typical filtering scale, because halos at the filtering scale of, e.g., Refs. \cite{1996ApJ...465..608T, 2000ApJ...542..535G, 2008MNRAS.390..920O} still have significant gas. We note that our definition of luminous fraction is a statement about whether there is \emph{any} star formation or not, so we err on the side of a lower filtering scale to allow for even a small amount of post-reionization star formation.  Note that our pre- and post-reionization thresholds are low compared to the literature (cf. \cite{Lunnan:2011cf, Bose:2018vpe}), so we predict more luminous satellites than other works.

One may translate this criterion on $v_\mathrm{max}$ to a luminous fraction as a function of M$_\mathrm{infall}$ given a set of halo merger trees, as was done in Ref. \cite{2017MNRAS.471.4894D}.  For the reionization redshifts considered in this work, the luminous fraction plummets at $M_\mathrm{infall} \approx 10^8$ M$_\odot$.  This is higher than but consistent with the 10$^7$ M$_\odot$ pre-reionization threshold for three reasons.  First, a halo of mass 10$^7$ M$_\odot$ at $z=10$ is typically much more massive (by an order of magnitude) before infall on a MW-like host if it survives to the present.  This is true even if the halo only undergoes ``passive accretion,"
i.e., it only appears to grow on account of the reduced critical density at late times, which cause us to identify increasingly more mass with the halo \cite{diemer2013}.  Second, a 10$^7 M_\odot$ halo is still cosmically quite rare at $z>10$.  Third, and related to this last point, most halos with M$_\text{infall} = 10^8 M_\odot$ today were below the pre-reionization go condition.

Our choices are backed by simulations, but where ambiguity exists, our model errs on the side of predicting too many, rather than too few, luminous dwarfs.

\subsection{Redshift-dependence of the reionization model}

As described in the previous section and in Sec. III of the main text, feionization suppresses the formation of galaxies in small halos \cite{2000ApJ...539..517B}.  Our fiducial model, based on Ref. \cite{2017MNRAS.471.4894D}, assumed that the reionization of the Local Group occurred at $z_\text{re} = 9.3$, at the early end for global reionization \cite{2016A&A...596A.108P}.  However, reionization is expected to be patchy.  Depending on the exact nature of the source of reionizing photons at high redshift, the MW is expected to reionize earlier than average \cite{2017arXiv170306140D,Aubert:2018cqm}.  Fewer halos form stars if reionization occurs earlier.

\begin{figure}[t!]
\includegraphics[width=0.5\textwidth]{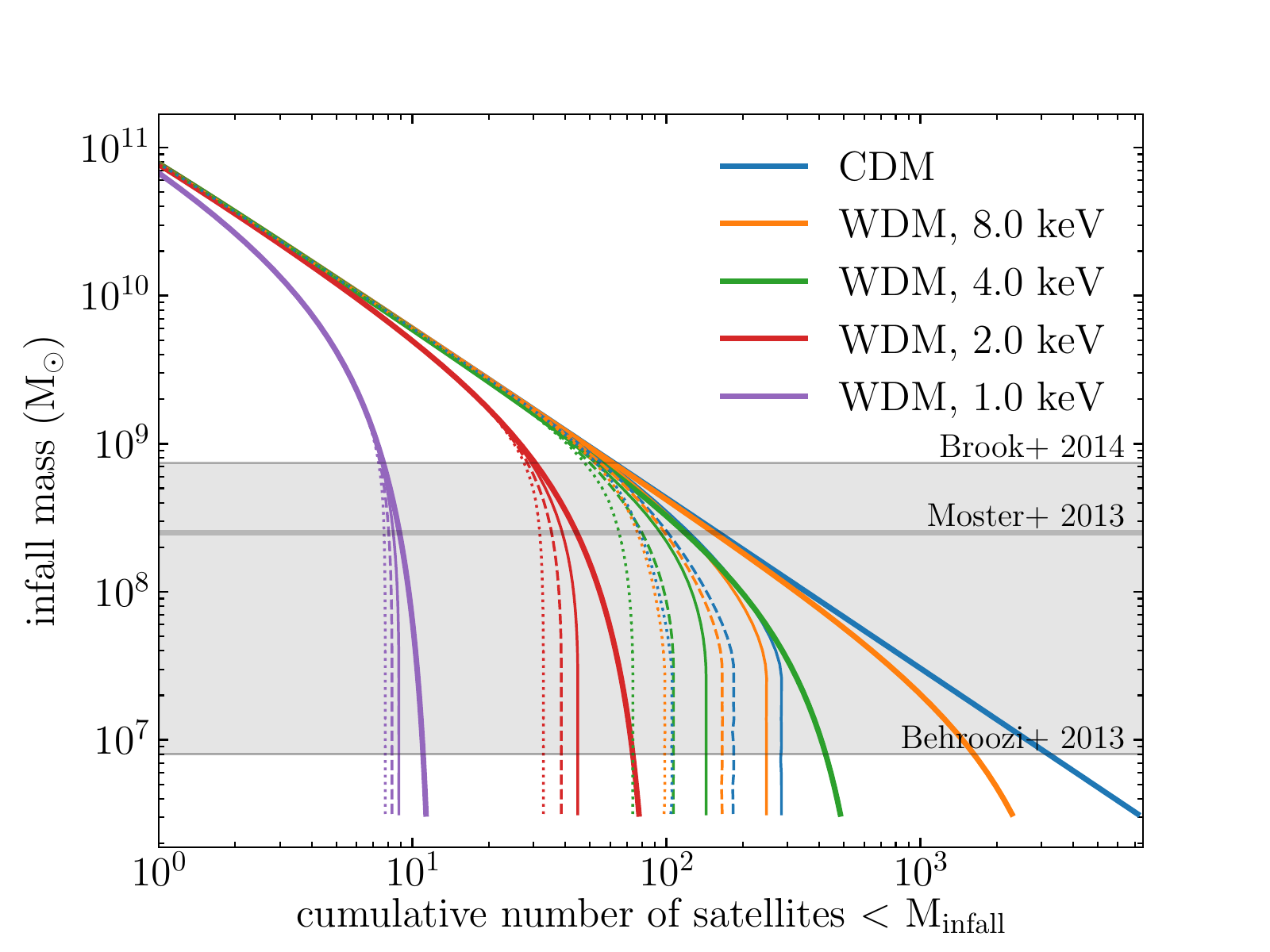}
\caption{Similar to Fig. 3, but with additional reionization redshifts.  Shown are the subhalo mass functions (solid bold line) and galaxy mass functions assuming $z_\text{re}$ = 9.3 (thin solid line), 11.3 (dashed) and 14.4 (dotted).}
\label{fig:mf_nsubs-allz}
\end{figure}

\begin{figure*}
\includegraphics[width=0.49\textwidth]{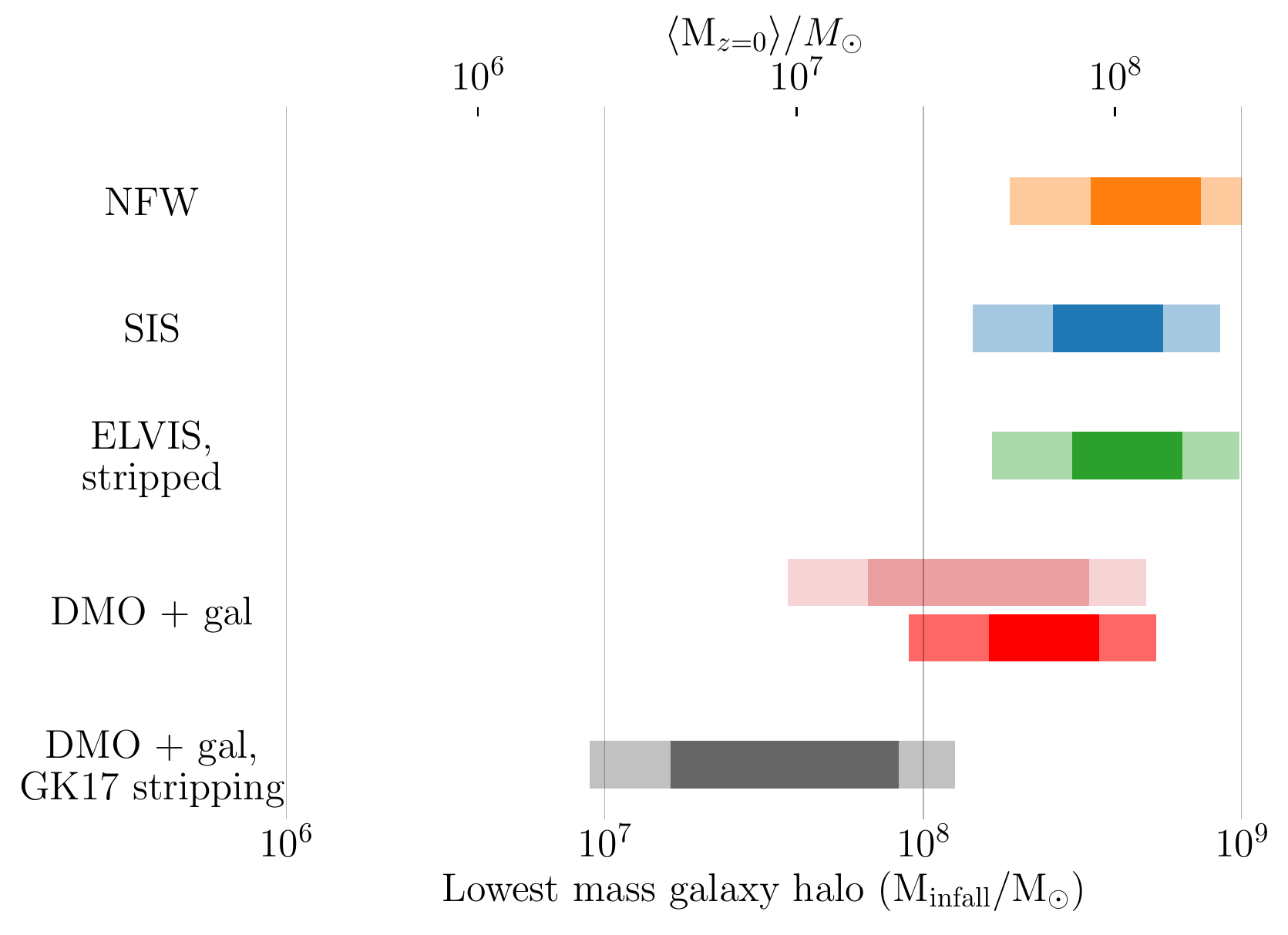}
\includegraphics[width=0.49\textwidth]{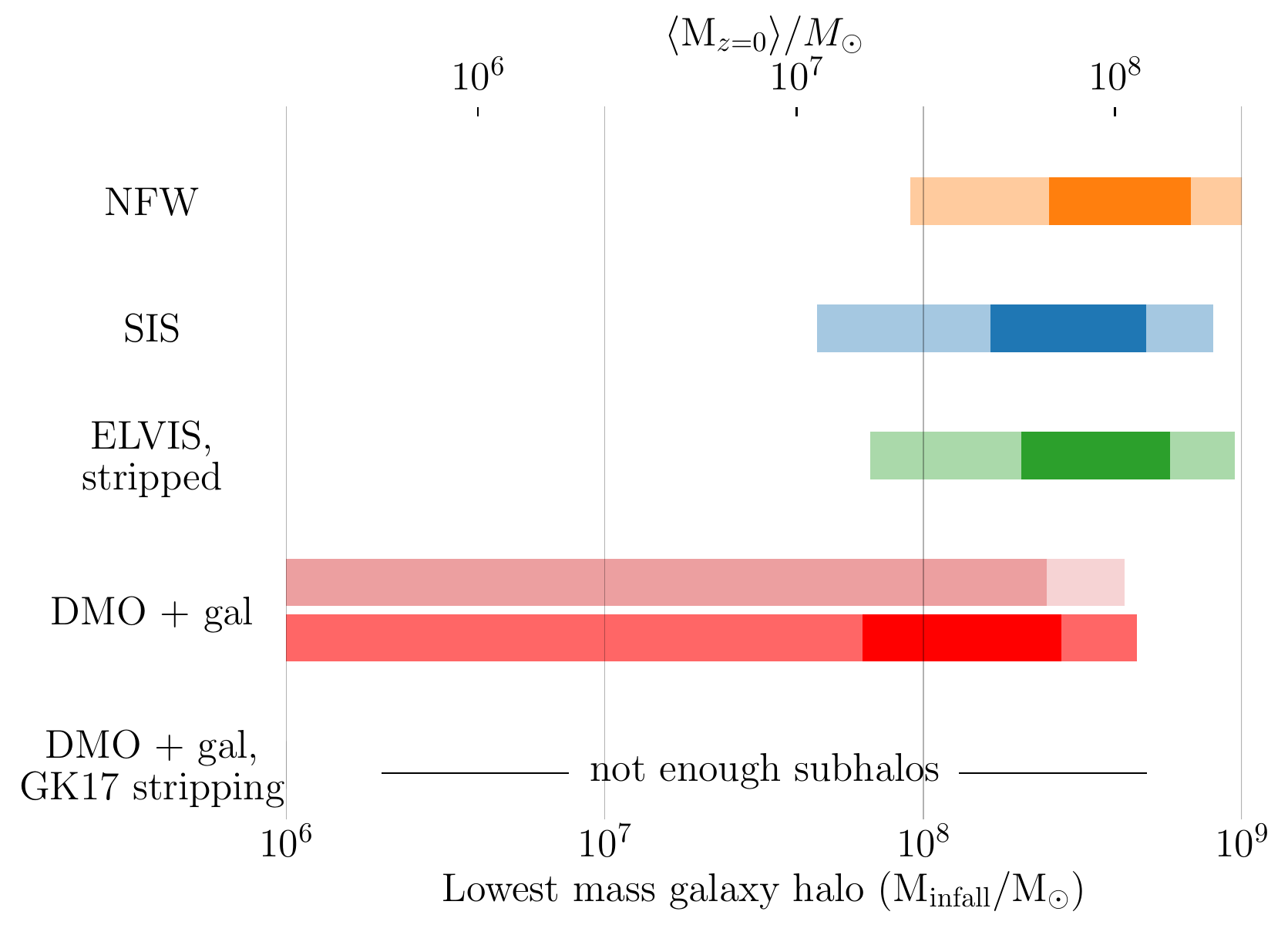} \\
\includegraphics[width=0.49\textwidth]{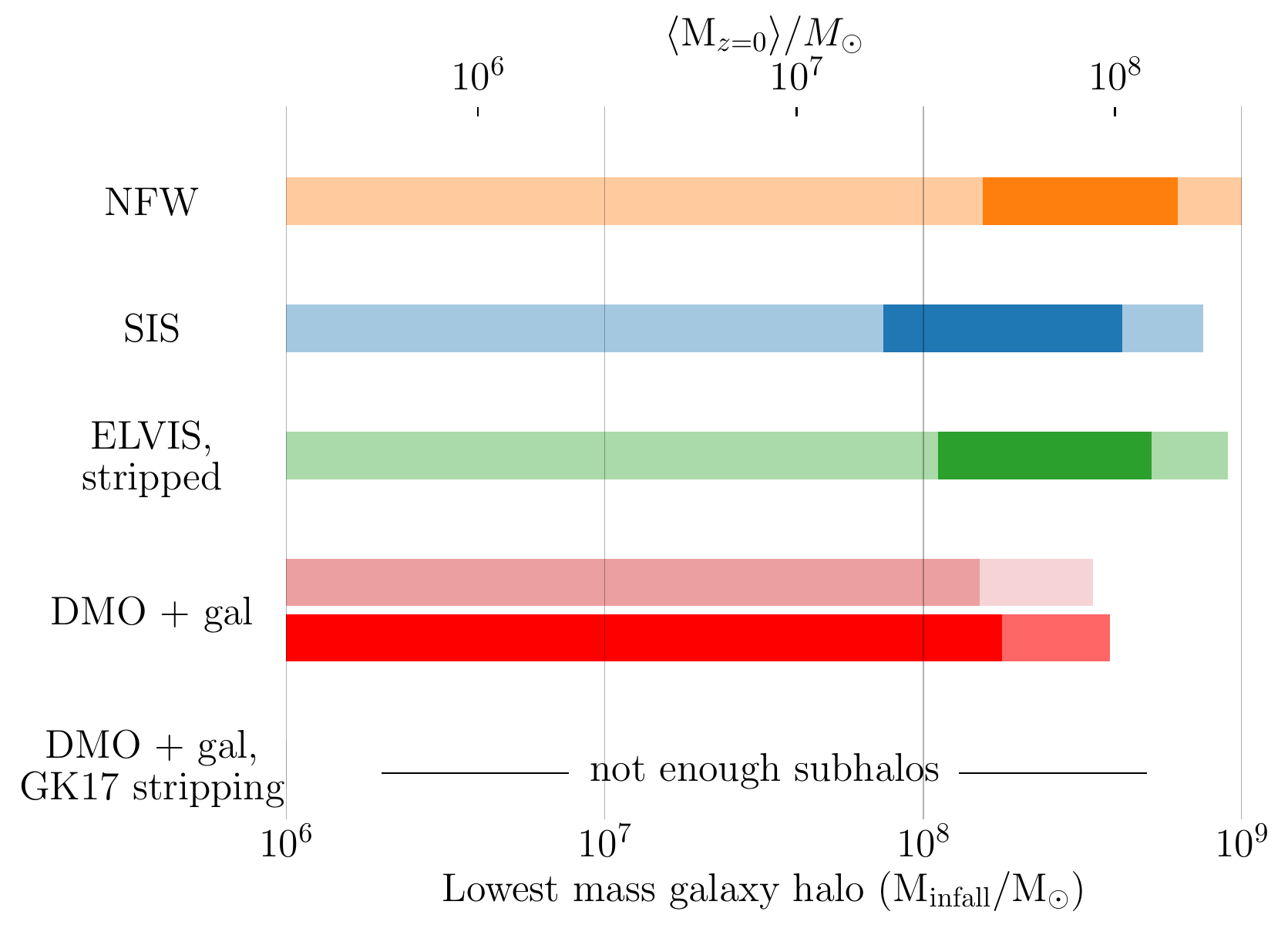}
\includegraphics[width=0.49\textwidth]{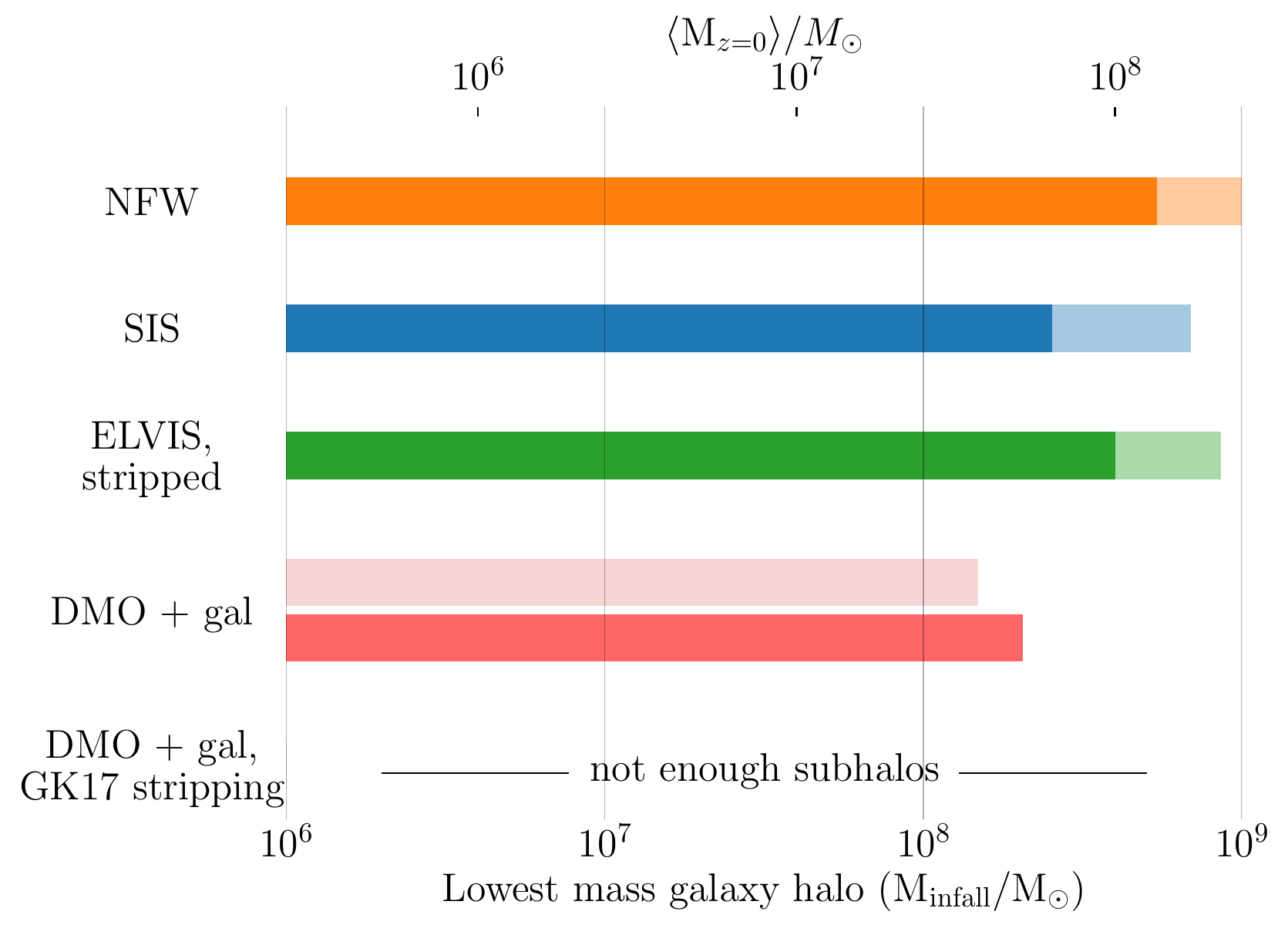}
\caption{Similar to the right panel of Fig. 2, but for different reionization redshifts.  The mass of faint galaxies similar in luminosity to Segue I, assuming all subhalos can host a galaxy (i.e. no reionization, top left), or a reionization redshift of $z_\text{re}$ = 9.3 (right).  Bottom row: $z_\text{re}$ = 11.3 (left) and 14.4 (right).}
\label{fig:min_mass-allz}
\end{figure*}

In Fig. \ref{fig:mf_nsubs-allz} and Tab. \ref{tab:mf_nsubs}, we show the maximum number of galaxies that form (i.e. integrating down to the threshold mass required to form luminous galaxy), assuming reionization occurs at earlier redshifts of $z_\text{re}$ = 11.3 and 14.4 for CDM as well as WDM models.  Note that should one integrate down to e.g. Segue I's halo mass, or any mass above this threshold, the number of galaxies will be strictly lower than the values listed in in Table \ref{tab:mf_nsubs}.  For comparison, we show the total number of subhalos.  Assuming the Milky Way has a completeness-corrected total of $\sim$120-150 luminous galaxies brighter than Segue I, there is tension with even the CDM predictions for $z_\text{re}$ = 11.  Reionization redshifts $z_\text{re} \gtrsim 11$ are disfavored.  An early reionization redshift also puts severe pressure on WDM models.  In Fig. \ref{fig:min_mass-allz}, we show the implied minimum halo mass if we match the prediction for CDM with the completeness-corrected galaxy count.

\begin{table}[h]
\caption{Number of satellites and galaxies down to 10$^6$ M$_\odot$}
\label{tab:mf_nsubs}
\begin{tabular}[b]{l c c c c} \hline
 & \multirow{2}{*}{subhalos} & \multicolumn{3}{c}{galaxies} \\
 & & $z$ = 9.3 & $z$ = 11.3 & $z$ = 14.4 \\ \hline
CDM        & $\infty$ & 284 & 183 & 104 \\
WDM, 8 keV & 3030     & 248 & 166 & 98  \\
WDM, 4 keV & 540      & 143 & 106 & 74  \\
WDM, 2 keV & 82       & 45  & 39  & 33  \\
WDM, 1 keV & 12       & 9   & 8   & 8   \\ \hline
\end{tabular}
\end{table}

\subsection{Stellar mass of Segue I}\label{appendix:mstar}

In order to compare the expected number of super-Segue-I-mass satellites of the Milky Way to the the completeness-corrected satellite counts, we must determine the infall subhalo mass of Segue I.  As discussed in Sec. III, the first step in deriving this is to convert its luminosity into a stellar mass.  While the estimates of the stellar masses of the luminous classical satellites are complicated by their extended star-formation histories \citep{Woo:2008gg}, ultrafaint galaxies like Segue I are comparatively simple systems.  Because they are ancient and metal-poor, the main uncertainty in translating from light to mass, or mass to light, is the initial mass function (IMF) of stars down to the hydrogen burning limit.  

For the ultrafaints, we assumed a V-band mass-to-light ratio $M_*/L_V = 2 M_\odot/L_\odot$ for Segue I.  This estimate gives us a stellar mass of 680 M$_\odot$, close to that computed by \citep{martin2008}, who found Segue I to have $M_*$ = 600 M$_\odot$ assuming a Kroupa IMF.  Assuming instead a Salpeter IMF produces a larger estimate of M$_*$ = 1300 M$_\odot$.  Should this larger stellar mass be adopted, Segue I would have a higher halo mass, and the missing satellites problem would be further lessened in severity.  However, recent work on six ultrafaint dwarfs using deep HST photometry suggests that their IMFs are closer to the Kroupa than Salpeter IMF \citep{2018ApJ...855...20G}.  We consider $M_*/L_V = 2 M_\odot/L_\odot$, which lies between Kroupa and Salpeter, a reasonable estimate for this work.

\subsection{Matching galaxies to halos}

The final step in predicting the number of luminous satellites brighter than Segue I in the MW halo is to estimate Segue  I's subhalo mass from its stellar mass (inferred from its luminosity, see Sec. III and \ref{appendix:mstar}).  To do this, we use relations that relate the stellar mass to \emph{infall} halo mass, what are known as stellar mass-halo mass (SMHM) relations.

For galaxies more massive or luminous than the Fornax MW satellite, matching galaxies to halos before infall provides the best fit to observational galaxy clustering, abundance, and growth patterns \cite{2013ApJ...779...25B, 2013MNRAS.428.3121M, 2013ApJ...771...30R, 2018MNRAS.477..359C}.  However, no stellar mass-halo mass relation (SMHM) has been constructed empirically at lower stellar masses, the regime of the SDSS ultrafaints.  However, high-resolution hydrodynamic simulations of dwarf galaxies in field environments indicate that extrapolations of empirically well-calibrated relations at $M_*\sim 10^7M_\odot$ scales are likely valid down to at least a few $10^3 M_\odot$ in stellar mass \cite{2017arXiv170506286M, 2017arXiv170206148H, Ma:2017avo, 2018arXiv180106187F}.  In simulations, SMHM relations are in place before reionization \cite{2015ApJ...807L..12O}.

We used a one-to-one relation between stellar mass and halo mass (i.e. placed the brightest satellites in the most massive luminous halos and so on down the respective mass functions, \`a la abundance matching) to estimate satellite number counts.  However, there is certainly scatter in the relation, with some suggestions that the scatter increase for decreasing halo mass \cite{2017MNRAS.464.3108G}. Scatter reduces the precision with which we can identify Segue I's mass, and requires us to consider a range of possible masses.  We sought to capture this uncertainty by considering a broad range of SMHM relations, which spans more than an order of magnitude in halo mass.  Further, Ref. \cite{2016arXiv161207834J} calculated completeness corrections using SMHMs with and without scatter and found no differences.  Ref.  \cite{2017MNRAS.464.3108G} noted that the observed SMHM relation may be shallower than the true SMHM; this would map Segue I to a larger halo mass, causing the missing satellites problem to further lessen in severity.

As a further check, we calculated the number of luminous satellites above a fixed stellar mass threshold and a fixed subhalo mass function using a probabilistic model to capture the scatter,
\begin{equation}
\begin{aligned}
N(>M_*^\text{thresh}) &= \int_{\log M_*^\text{thresh}}^{\log M_*^\text{host}} d\log M_* \\
&\quad \times \int_{0}^{M_\text{vir}^\text{host}} dM~f_\mathrm{lum}(M)~\frac{dN}{dM}(M_{vir}^{host}) \\
&\quad \times P(\log M_* | \log M ),
\end{aligned}
\end{equation}
where $M$ is the infall subhalo mass, $M_\text{vir}^\text{host}$ and $M_*^\text{host}$ are the host virial and stellar masses, and $P(\log M_* | \log M )$ is a model for the scatter in the SMHM relation (the probability that a satellite has stellar mass $M_*$ given an infall halo mass $M$).  We take a log-normal distribution for $P(\log M_* | \log M )$, hence the logarithmic variables and integration limits.  Because the (sub)halo mass function falls so steeply for increasing stellar mass, scatter in the SMHM relation is more likely to push a low-mass halo above the $M_*$ threshold than vice versa.  We find that a 1 dex scatter increases the number of predicted satellites up by 50\% near the estimated Segue I stellar mass.  Higher scatter does not increase the predicted satellite count further because of the $f_{lum}$ cutoff at small halo mass.  The scatter-induced increase in the number of predicted satellites is smaller for higher stellar mass thresholds.  We note that the scatter-induced uncertainty in the abundance of satellites more massive than Segue I is subdominant to the uncertainty in the mean SMHM relation which was accounted for in the main text.

\subsection{Treatment of warm dark matter}

We adopt simple simulation-driven models to assess the missing satellites problem in the context of dark matter models with a truncation in the matter power spectrum.  Although we refer to these models as ``warm dark matter'' models here, and benchmark against a thermal relic treatment, our constraints can be applied to any dark matter model (e.g., sterile neutrinos, hidden-sector dark matter, even fuzzy dark matter) with a small-scale suppression in the matter power spectrum.

\vskip 0.1cm

\noindent
\emph{Completeness correction:} WDM simulations show that the radial distribution of subhalos is identical to CDM in dark-matter-only simulations \cite{2014MNRAS.439..300L}.  Massive luminous satellites in hydrodynamic WDM simulations share this similarity \cite{2017MNRAS.468.4285L}.  This fact has its origins in the effect of the truncation of the matter power spectrum on halo structural properties.  Only halos near the half-mode mass have halo properties that differ noticeably from CDM \cite{colin2015,Bose:2015mga}.  However, on this scale, the halo mass function is significantly suppressed \cite{Maccio:2012qf}.  Thus, halos above the half-mode mass should behave as CDM halos.  In hydrodynamic simulations, the central density profiles of satellites are driven by baryons in both CDM and WDM cosmologies.  Therefore, we do not expect significant differences between WDM and CDM satellite tidal stripping histories, and hence the radial distribution of satellites should look similar.  Thus, the completeness corrections we found in the main text apply to WDM cosmologies as well as to CDM.

One caveat to this argument pertains to the small number of satellites that form in halos below the half-mode mass.  These halos form late and with low densities relative to their CDM counterparts \cite{Schneider:2014rda}.  Such halos are more prone to tidal disruption; additionally, their late formation and infall onto the host implies that they are not centrally concentrated in the host (see, e.g., Ref. \cite{Rocha:2011aa}).  The combination of these effects suggests that our less centrally concentrated completeness corrections may apply for some of the observed ultrafaint galaxies.  This goes in the direction of exacerbating the ``too many satellites'' problem, and leads to more stringent constraints on the matter power spectrum cut-off.

\vskip 0.1cm
\noindent
\emph{Satellite number count predictions:} Our prediction for the number of luminous satellites in WDM cosmology mirrors our CDM treatment, although with a few crucial caveats.

First, we consider the subhalo mass function, the scaffolding for our satellite prediction.  As with CDM, this requires the mass function of surviving subhalos (again defining mass to be the mass at infall).  Ref. \cite{2012MNRAS.424..684S} find that the WDM halo mass function for field halos is suppressed relative to CDM according to
\begin{eqnarray}
 \frac{n_{WDM}}{n_{CDM}} = \left( 1 + M_\text{hm}/M \right)^{-\beta}
\end{eqnarray}
for $n$'s representing the number of halos per unit volume, a half-mode mass $M_\text{hm}$, a virial mass $M$, and $\beta = 1.16$.  Assuming most subhalos were originally field halos that fell into a bigger halo, we apply this suppression function to the CDM subhalo infall mass function.  We checked the resulting number counts against other work (e.g., Ref. \cite{Bose:2015mga}) and found consistent results.

Second, we consider the luminous fraction.  Conservatively, we assume that the luminous fraction is the same in WDM as in CDM.  This assumption is conservative in the sense that it almost certainly predicts too many luminous satellites for fixed halo mass, especially at the low-mass end.  Halos, and hence their luminous components, form later in WDM cosmologies than in CDM \cite{Governato:2014gja, 2018arXiv180305424B}.  Halos below the half-mass scale form much later \cite{Schneider:2014rda}.  This further suppresses the number of luminous galaxies in WDM.  In addition, although WDM models can produce enough moderate-mass galaxies to reionize the universe at a time consistent with observations \cite{2016MNRAS.463.3848B}, they do so in bigger halos that form stars more rapidly than in CDM.  After reionization, star formation is suppressed in low mass WDM halos in the same way as in CDM.  This also suggests that the luminous fraction of $\sim 10^8 M_\odot$ halos is further suppressed in WDM than CDM cosmologies.  Our simple model thus predicts more luminous WDM satellites than we should realistically expect.  Once luminous fractions are robustly obtained from simulations and semi-analytic models, they can be used in our model for simple satellite predictions.

Finally, we consider the SMHM in WDM.  Above the half-mode mass scale, where hydrodynamic simulations have been performed and where some semi-analytic models exist, the SMHM relations looks consistent with CDM, albeit with perhaps somewhat more scatter \cite{2017MNRAS.468.4285L, 2018arXiv180305424B}.  Thus, we use CDM SMHM relations for our WDM predictions.

In conclusion, we made a set of conservative choices to compare completeness-corrected satellite counts to theoretical satellite predictions for WDM.  We consider it likely that true constraints are more stringent than what we show in this work.


\section{Luminosity Functions}

\begin{figure*}
\centerline{\includegraphics[width=0.5\textwidth]{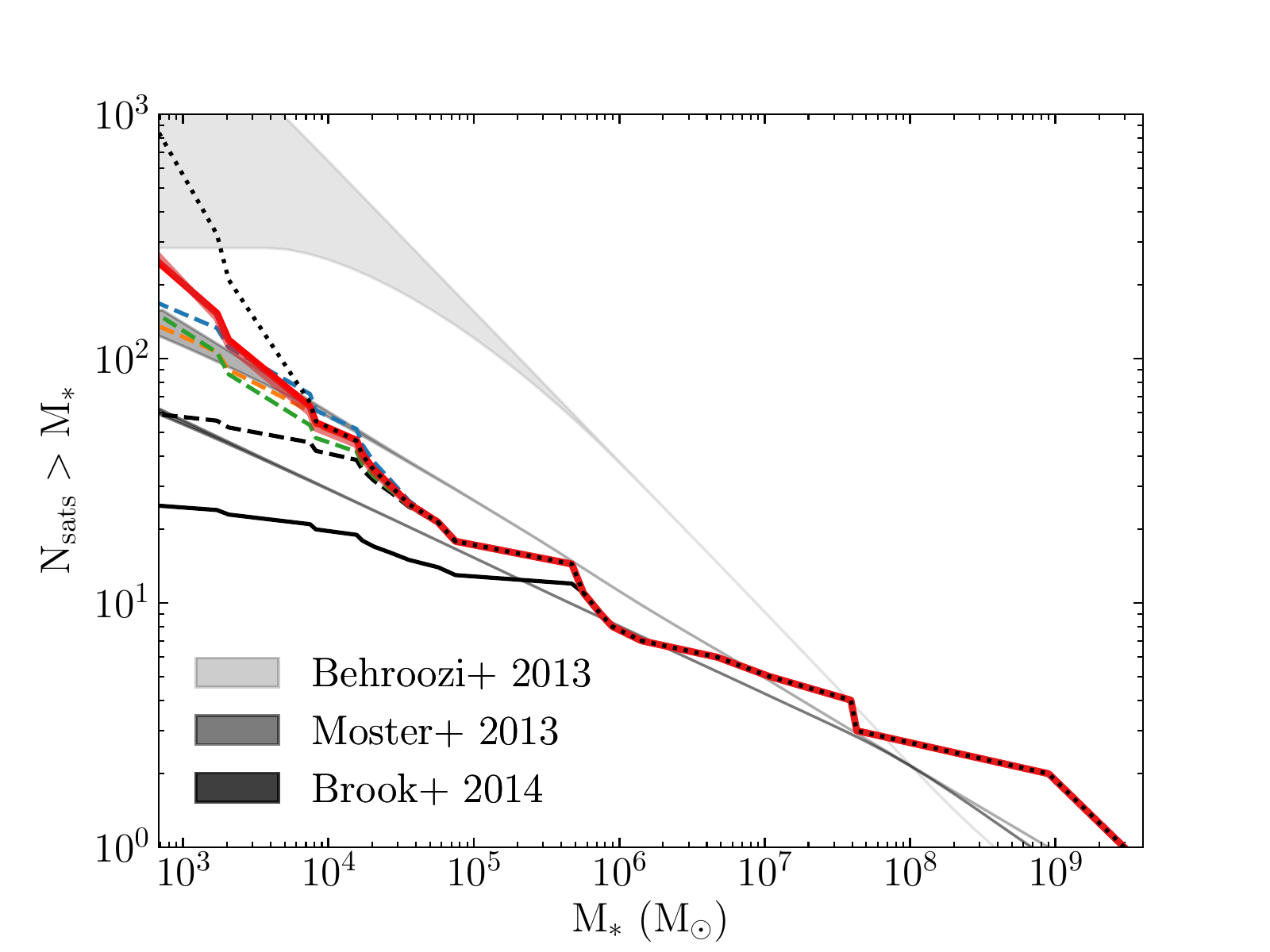}
\includegraphics[width=0.5\textwidth]{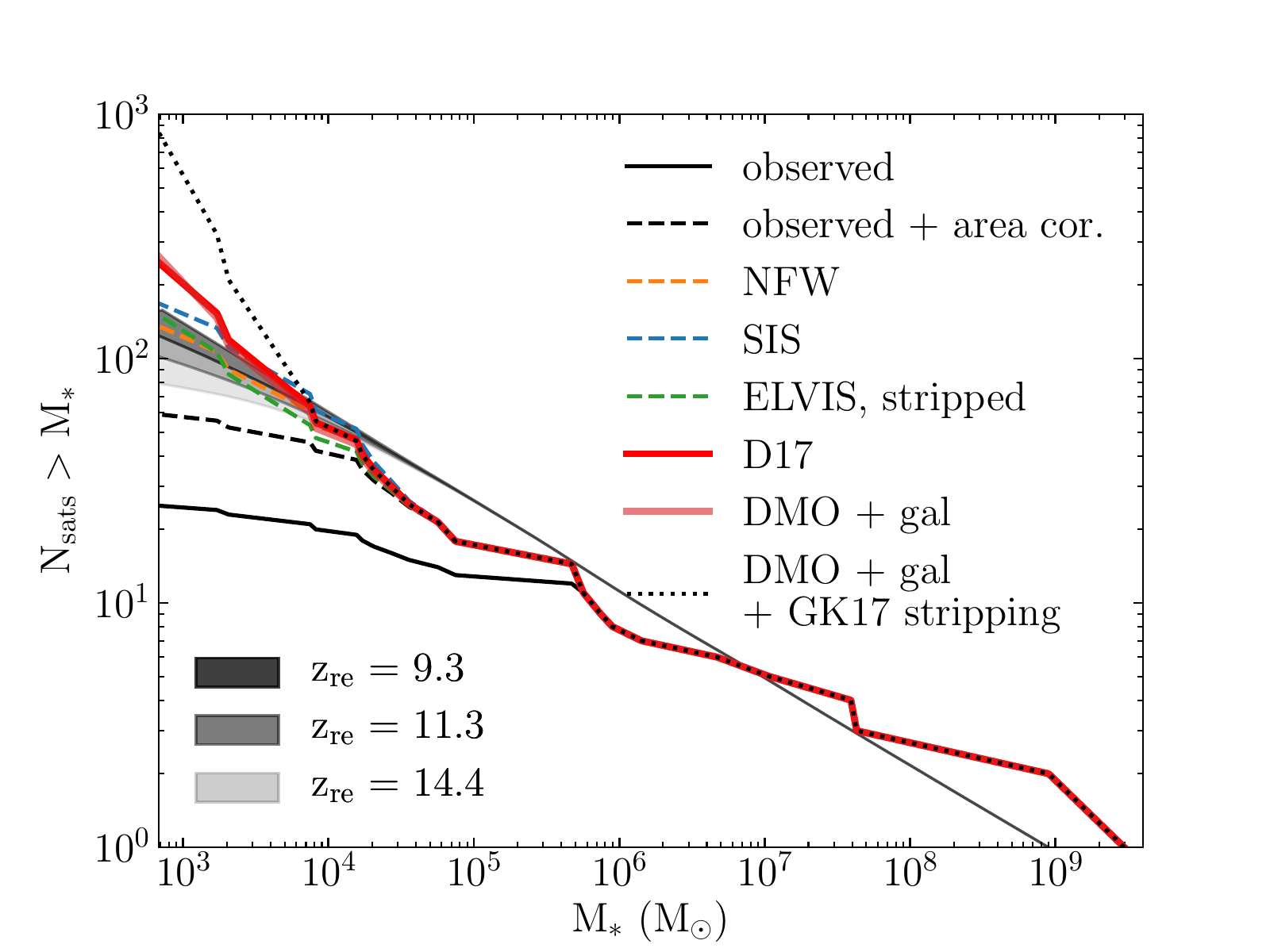}}
\caption{Completeness corrected luminosity functions.  Uncertainties (e.g. from anisotropies) have not been plotted.  The uncorrected (black) and only area-corrected (i.e. no radial correction; black dashed) luminosity functions are also shown.  \emph{Left}: The gray-shaded bands represent the luminosity functions predicted by different SMHM relations (see Appendix B for details).  The lower edge of each band is generated by assuming a reionization redshift of $z_\text{re}$ = 9.3, while the upper edge assumes that all subhalos host a galaxy.  Note that the Brook et al. 2014 (Ref. \cite{2014ApJ...784L..14B}) relation predicts far fewer and the Behroozi et al. 2013 (Ref. \cite{2013ApJ...770...57B}) relation predicts far more than the completeness corrected counts.  \emph{Right}: Same as left, except the gray-shaded bands assume a Moster et al. 2013 (Ref. \cite{2013MNRAS.428.3121M}) SMHM relation, and instead $z_\text{re}$ is varied from 9.3 to 14.4.  The centrally concentrated radial distributions derived from dark matter only simulations produce the smallest number of satellites, about 125-150, at M$_V$ = -1.5 (L$ = 340 $L$_\odot$).  Those from baryonic simulations with tidal stripping, which are not centrally concentrated, have the largest corrections.}
\label{fig:lumfxn}
\end{figure*}

The completeness corrections derived in Sec. II can also be used to estimate luminosity functions.  In Fig. \ref{fig:lumfxn}, we show the luminosity function of MW satellites within 300 kpc of the MW center for the radial distributions discussed in Sec. II.  The luminosity function is plotted both in units of solar luminosity and in the V-band absolute magnitude M$_\text{V}$ convention of the field.  In comparison, we have also plotted the luminosity functions predicted by CDM, assuming the CDM mass function, the luminous fraction, and the SMHMs considered in Sec. III.  These are plotted with gray bands.  The width of the bands denotes the uncertainty in the reionization redshift; the lower limit of each band is derived by assuming the fraction luminous if reionization occurs at $z_\text{re} = 9.3$, while the upper limit assumes all subhalos are luminous.

A remarkable conclusion is that the predicted luminosity function with our fiducial model, with the SMHM of  \cite{2013MNRAS.428.3121M} (labeled Moster et al. in Fig. \ref{fig:lumfxn}) is in detail a good fit to our fiducial completeness-corrected luminosity function.  Note that the Brook et al. SMHM relation underestimates the number of completeness-corrected satellites, predicting only as many satellites as inferred from a simple area correction; it cannot reproduce the number of satellites inferred when the radial correction is taken into account.  In fact, it underpredicts the number of \emph{known} satellites.  The Behroozi et al. SMHM relation predicts far more satellites than inferred from the completeness corrections, requiring us to place galaxies in extremely low mass halos as low as 10$^6$ M$_\odot$.  Even for Moster et al., however, there is a mismatch in the completeness-corrected and theoretical luminosity functions at stellar masses of about 10$^{4-5}$ M$_\odot$.  This dip has been noted by D17 and Ref. \cite{Bose:2018vpe}, and can be potentially explained by tidal stripping \cite{Brooks:2012ah}.


\section{Removing Segue I}

The closest and faintest satellites dominate the completeness correction.  We re-ran our analysis without Segue I, the faintest dwarf in our sample (L$=340$L$_\odot$, or $M_V$ = -1.5) and located $~$23 kpc from the Sun.  For our fiducial radial distribution (D17), the corrected satellite count for the full sky drops by 40\% from 236 to 142, for satellites with luminosity $L \ge 850$ L$_\odot$.  Counts for the other distributions are listed in Tab. \ref{tab:counts-noSegI}.

\begin{table}
\caption{Completeness corrected satellite (L$>850$L$_\odot$) counts without Segue I}
\label{tab:counts-noSegI}
\begin{tabular}[t]{l c c c c} \hline
 & & \multicolumn{3}{c}{Predictions} \\
distribution & $r_\text{1/2}$ & all sky & DES & LSST Year 1 \\ \hline
NFW             & 124 kpc     & 95      & 9     & 34     \\
SIS             & 150 kpc     & 122     & 11    & 38     \\
ELVIS, stripped & 90 kpc      & 94      & 10    & 37     \\
D17             & 124 kpc     & 142     & 13    & 44     \\
sims            & 110-158 kpc & 135-258 & 13-21 & 44-58  \\
sims + GK17     & 130-170 kpc & 130-638 & 28-44 & 77-100 \\ \hline
\end{tabular}
\end{table}

\end{document}